\documentclass[pra,showpacs,twocolumn,superscriptaddress]{revtex4}
\newcommand{\ket}[1]{|{#1}\rangle}

\newcommand{\be}{\begin{equation}}
\newcommand{\ee}{\end{equation}}

\usepackage{amsfonts}
\usepackage{amsmath}
\usepackage{graphicx}
\usepackage{amssymb}
\usepackage{amsmath}
\usepackage{amssymb}
\usepackage{graphicx}
\usepackage{lscape}
\usepackage{color}
\usepackage{epstopdf}

\begin{document}
\title{Super Hamiltonian in superspace for incommensurate superlattices and quasicrystals}

\author{M. Valiente}\email{mvaliente@tsinghua.edu.cn}
\affiliation{Institute for Advanced Study, Tsinghua University, Beijing 100084, China}
\author{C.~W. Duncan}
\affiliation{SUPA, Institute of Photonics and Quantum Sciences, Heriot-Watt University, Edinburgh EH14 4AS, United Kingdom}
\author{N.~T. Zinner}
\affiliation{Department of Physics \& Astronomy, Aarhus University, Ny Munkegade 120, 8000 Aarhus C, Denmark}
\affiliation{Aarhus Institute of Advanced Studies, Aarhus University, H{\o}egh-Guldbergs Gade 6B, 8000 Aarhus C, Denmark}

\begin{abstract}
Infinite quasiperiodic arrangements in space, such as quasicrystals, are typically described as projections of higher-dimensional periodic lattices onto the physical dimension. The concept of a reference higher-dimensional space, called a superspace, has proved useful in relation to quasiperiodic systems. Although some quantum-mechanical systems in quasiperiodic media have been shown to admit quasiperiodic states, any sort of general Hamiltonian formalism in superspace is lacking to this date. Here, we show how to extend generic quantum-mechanical Hamiltonians to higher dimensions in such a way that eigenstates of the original Hamiltonian are obtained as projections of the Hamiltonian in superspace, which we call the super Hamiltonian. We apply the super Hamiltonian formalism to a simple, yet realistic one-dimensional quantum particle in a quasiperiodic potential without the tight-binding approximation, and obtain continuously labelled eigenstates of the system corresponding to a continuous spectrum. All states corresponding to the continuum are quasiperiodic. We also obtain the Green's functions for continuum states in closed form and, from them, the density of states and local density of states, and scattering states off defects and impurities. The closed form of this one-dimensional Green's function is equally valid for any continuum state in any one-dimensional single-particle quantum system admitting continuous spectrum. With the basis set we use, which is periodic in superspace, and therefore quasiperiodic in physical space, we find that Anderson-localised states are also quasiperiodic if distributional solutions are admitted, but circumvent this difficulty by generalising the superspace method to open boundary conditions. We also obtain an accurate estimate of the critical point where the ground state of the system changes from delocalised to Anderson localised, and of the critical exponent for the effective mass. Finally, we calculate, within the superspace formalism, topological edge states for the semi-infinite system, and observe that these exist, in the delocalised phase, within all spectral gaps we have been able to resolve. Our formalism opens up a plethora of possibilities for studying the physics of electrons, atoms or light in quasicrystalline and other aperiodic media.          
\end{abstract}
\pacs{
}
\maketitle
\section{Introduction}
Quasiperiodic and quasicrystalline structures in space have fascinated scientists for many decades, even before the experimental discovery of quasicrystals by Shechtman and co-workers in 1984 \cite{Shechtman}. In their work, Shechtman {\it et al.} observed a solid with symmetry forbidden for a crystal lattice, yet with long-range orientational order. Electron diffraction experiments showed a discrete, highly peaked pattern that could not correspond to a Bravais lattice. This type of arrangement was investigated theoretically soon after and the term ``quasicrystal'' was coined by Levine and Steinhardt in Ref.~\cite{LevineSteinhardt}. There, the authors (i) gave a crystallographic, Fourier-based definition of quasicrystals as arrangements of points whose diffraction pattern consists of a self-similar arrangement of infinitely sharp Bragg peaks -- just as for ordinary crystals -- in reciprocal space; and (ii) showed that the set of points conforming a quasicrystalline spatial arrangement is spanned by a number of lattice vectors which exceeds the spatial dimension of the quasicrystal. These important properties had been known in the mathematical literature, from de Bruijn's work \cite{deBruijn} on Penrose's tilings \cite{Penrose}. From these, it is inferred that non-unique extensions to dimensions higher than the physical dimension are useful in constructing quasiperiodic arrangements of points, i.e. quasicrystalline structures. This fact resulted in the development of cut-and-project methods \cite{CutProject}, and what is now commonly known as superspace methods \cite{Elcoro}. A related, yet rather different topic is the theory of quasiperiodic and almost periodic functions, their foundations dating back to at least 1925 with the seminal work of Harald Bohr \cite{Bohr}. In short, a quasiperiodic function with a $d$-dimensional domain is defined as a projection of a function with a higher dimensional domain -- superspace -- to $d$ dimensions, in a way that will be specified below (see section \ref{superspace}).

The above formal properties of quasicrystals and quasiperiodic functions would be mere curiosities if they did not result in interesting physical consequences. Fortunately, there is a plethora of unusual phenomena associated with quantum-mechanical systems in quasiperiodic media, and of the structural physical components of the quasicrystals themselves. Not long after the discovery of quasicrystals, a hydrodynamic description of the quasiperiodic lattice showed the existence of phason modes \cite{Bak,Levine2}, which are deeply connected with the underlying concept of superspace \cite{CutProject,Kalugin,Duneau,PhasonsReview}. Regarding their electronic properties, pseudogaps, with a significantly reduced density of states near the Fermi energy, have been predicted \cite{Hafner} and observed \cite{Lynch}. Quasicrystal structures may also be artificially engineered using photonic lattices \cite{Rechtsman2}. For instance, disorder has been experimentally shown to enhance transport of light by coupling it to states in a pseudogap using photonic quasicrystals \cite{Rechtsman1}, as proposed theoretically for electronic systems decades ago \cite{Fujiwara}. There is also increasing interest in incommensurate systems, an interesting example of it being superconductivity in quasiperiodic host-guest structures \cite{Brown}.  Very recently, the first quasicrystalline optical lattice for ultracold neutral atoms was realized by Schneider and co-workers \cite{Schneider}, opening the door to the possible investigation of strongly correlated physics in a quasicrystalline potential \cite{Johnstone}. In Ref.~\cite{Schneider}, the quasicrystal structure is two-dimensional and has an 8-fold rotational symmetry, generated by two copies of a square lattice rotated 45 degrees with respect to each other. A Bose-Einstein condensate (BEC) of $^{39}\mathrm{K}$ atoms was prepared in a trap and rendered essentially non-interacting by means of a magnetic Feshbach resonance \cite{Cachibache}, and was subsequently released, thereby scattering off the two-dimensional optical lattice. This experiment showed that, at short times, the BEC behaves as if it performed a quantum walk on a four-dimensional square lattice, yet another manifestation of the relevance of the concept of superspace.

Electrons, ultracold atoms or light in one-dimensional quasicrystals, i.e. subjected to a quasiperiodic potential or its analog in photonic lattices, also exhibit highly non-trivial phenomenology. Deep in the tight-binding regime, these systems may be described by the Aubry-Andre model \cite{Aubry}, which exhibits a transition from extended to Anderson localised eigenstates  \cite{Anderson} without the need of disorder as the strength of the quasiperiodic external potential approaches a critical point. This model, however, is unrealistic since all eigenstates are either localised, delocalised or critical for a given strength of the quasiperiodic potential. More realistic models, still within a tight-binding approach, show a so-called ``mobility-edge'' \cite{DasSarma}, i.e., eigenstates in some regions of the spectrum are localised while others are not. The existence of a mobility edge in a realistic system, i.e. without the tight-binding approximation, however, is utterly trivial, since for sufficiently high energies the spectrum is certainly continuous. In continuum models, moreover, it is known mathematically that localisation occurs, as the strength of the potential increases, first in the ground state and at low energies \cite{Frolich}, a fact that has been accurately quantified numerically in a recent article \cite{SanchezPalencia}. This means that we are generally safe to try and calculate low-energy continuous spectra as long as the ground state of the system remains delocalised.  

Combining the two similar concepts of superspace, that is, crystallographic and functional, in order to describe the quantum mechanics of particles in quasicrystal potentials, is a very appealing idea. In the mathematical literature, there are examples which show that generalised Bloch states, with quasiperiodic Bloch functions, can be rigorously defined as eigenfunctions of quasiperiodic Hamiltonians \cite{MoserRotationNumber,AvronSimon}. Given that quasiperiodic functions are defined as projections from functions in superspace, one immediately wonders whether it is possible, in physically relevant cases, to define a Hamiltonian theory in higher-dimensional space in such a way that (at least some of) the projections of the eigenstates back onto physical space constitute eigenstates of the original problem. Here, we develop such Hamiltonian formalism for continuous models of quasicrystals. To test our results, we apply the method to the simplest non-trivial quasiperiodic model in the continuum, i.e. without the tight-binding approximation, consisting of a single particle in a one-dimensional quasiperiodic potential consisting of the sum of two periodic potentials with incommensurate periods. We find that all delocalised states of the system, whose energies lie in the continuous spectrum, are not only of the generalised Bloch form, but are also quasiperiodic. By carefully assessing our results, we are able to assign a unique continuous label to every continuum state with dimensions of momentum, from which the density of states or the effective mass near the ground state can be calculated. This shows that this label corresponds to the so-called rotation number, which plays a fundamental role in spectral theory of quasiperiodic Schr{\"o}dinger operators \cite{MoserRotationNumber,AvronSimon}. We also obtain all the Green's functions in closed form, depending only on the (right-moving) eigenstate at a given energy in the continuous spectrum. Incidentally, the Green's functions that we obtain -- and all subsequent results that derive from them -- are the exact Green's functions for {\it any} state belonging to continuous spectrum in {\it any} single-particle model in one dimension, which is a very strong and general result. From the exact Green's function we can easily write down scattering states off defects and external finite-range potentials, which have important applications in quantum transport \cite{QuantumTransport}. The local density of states is obtained from the diagonal part of the Green's function and is a quasiperiodic quantity, which greatly simplifies the calculation of the density of states via what we call the local super density of states, which is nothing but the extension of the local density of states to superspace, where it is periodic. The results obtained for the density of states using the energy as a function of the continuous momentum label, or rotation number, are compared to those obtained via the local density of states, and are in perfect agreement. Since we have access to the effective mass near the ground state, we also study its delocalisation-localisation transition, obtaining an accurate estimate of the critical point for the same case studied in Ref.~\cite{SanchezPalencia} from a different perspective, and the critical exponent. The superspace approach, which is used to directly study the infinite-size limit of the system, requires a finite basis of periodic functions, and we find that in this way, convergence is not achieved for localised states. We show that, even though the solutions may still be periodic in superspace, they may be singular (i.e. distributional), and this is proved in the strong-coupling limit. Fortunately, it is still possible to generalise the superspace approach to study finite-length systems with open boundary conditions, which works equally well in the localised and delocalised regimes. We finally show how topological edge states are obtained for a semi-infinite system as exponentially decaying (or increasing) generalised quasi-perodic Bloch states, and propose a definition of the one-dimensional topological invariant -- the Zak's phase -- for strictly quasiperiodic systems without invoking periodic approximants.


\section{Superspace}\label{superspace}
A well-known concept in the theory of quasiperiodic functions is that of superspace \cite{Elcoro,Wolff,Janner,Janssen}. By definition, a quasiperiodic function in $\mathbb{R}^d$ is a function $f:\mathbb{R}^d\to \mathbb{C}$ such that there exists a periodic function $g:\mathbb{R}^{d}\times \mathbb{R}^{d'}\to \mathbb{C}$, with $d'=nd$, $n\in \mathbb{Z}_+$, with
\begin{equation}
  f(\mathbf{r}) = g(\mathbf{r},\mathbf{r},\ldots,\mathbf{r}).
\end{equation}
Our first goal is to establish a method to obtain eigenfunctions of quasiperiodic Hamiltonians by solving an equivalent problem in a higher dimensional superspace. In other words, we ask whether it is possible to build a Hamiltonian in superspace which is periodic and whose eigenstates, upon projection onto the original space, are eigenstates of the original quasiperiodic Hamiltonian. Although this is {\it a priori} not necessary, we shall restrict the functions in superspace to $\mathcal{H}_S=L^2(\mathbb{R}^{d+d'})$, with its usual scalar product. Since $\mathcal{H}_S$ is a separable Hilbert space, it admits a numerable orthonomal basis. Moreover, the basis can be chosen to be of the form
\begin{equation}
  B_S = \left\{\prod_{i=1}^{d+d'}\psi_{n_i}(x_i)\right\}_{\mathbf{n}\in \mathbb{Z}^{d+d'}},\label{basis}
\end{equation}
which is clearly the case since, for instance, we may choose harmonic oscillator eigenstates in each of the coordinates. Since any wave function in $\mathcal{H}_S$ can be expanded as a series in the elements of $B_S$, all we need to know is how the Hamiltonian in superspace -- which we shall call the super Hamiltonian -- must act on basis functions of the product type, in order to produce eigenstates of the original Hamiltonian after projection. Potential terms simply multiply basis functions, and we shall only briefly discuss them below. However, kinetic energy terms need to be handled with care. Let $H_0$ be the original kinetic energy operator, and $H_{0,S}$ the part of the super Hamiltonian that contains partial derivatives. Then, $H_{0,S}$ must satisfy
\begin{equation}
  \lim_{\mathbf{r}_1\to \mathbf{r}}\ldots\lim_{\mathbf{r}_n\to \mathbf{r}}H_{0,S}\prod_{j=0}^{n}\psi_{j}(\mathbf{r}_j) = -\frac{\hbar^2}{2m}\nabla_{\mathbf{r}}^2\left(\prod_{j=0}^{n}\psi_{j}(\mathbf{r})\right),\label{SuperHamiltonianCondition}
\end{equation}
where we have identified $\mathbf{r}_0\equiv \mathbf{r}$, and $d'=nd$. Note that condition (\ref{SuperHamiltonianCondition}) is nothing but the statement that the operation of the (free particle) super Hamiltonian on a wave function of the product type must coincide with the action of the Hamiltonian on the same wave function upon projection onto physical space. Since any eigenfunction can be expanded in a basis of this type, Eq.~(\ref{SuperHamiltonianCondition}) must be satisfied. Achieving condition (\ref{SuperHamiltonianCondition}) is quite simple, by beginning with the r.h.s. of Eq.~(\ref{SuperHamiltonianCondition}). The gradient of the product of $n+1$ functions satisfies Leibniz's rule:
\begin{equation}
  \nabla_{\mathbf{r}}\left(\prod_{j=0}^{n}\psi_{j}(\mathbf{r})\right)=\sum_{i=0}^{n}\nabla_{\mathbf{r}}\psi_i(\mathbf{r})\prod_{j=0,j\ne i}^{n} \psi_j(\mathbf{r}).
\end{equation}
In order to have an operator $\tilde{\nabla}_S$ in superspace whose projection back onto regular space satisfies Leibniz's rule above, we can simply write
\begin{equation}
  \tilde{\nabla}_S =\sum_{j=0}^{n}\nabla_{\mathbf{r}_j}.\label{superspacegradient}
\end{equation}
We note that the expression in Eq.~(\ref{superspacegradient}) for the gradient $\tilde{\nabla}_S$ in superspace, has been derived using functions of the product form. This applies for wave functions that are not of the product form, if they admit an expansion in a basis of the form of Eq.~(\ref{basis}), which is the case for smooth functions in superspace. This is illustrated in Appendix \ref{apendicitis2}. It is worth clarifying that, in Eq.~(\ref{superspacegradient}) and throughout the manuscript, the notation for the gradients is taken in the following sense
\begin{equation}
  \nabla_{\mathbf{r}_j} =\hat{e}_x\partial_{x_j}+\hat{e}_y\partial_{y_j}+\hat{e}_z\partial_{z_j},
\end{equation}
where $\hat{e}_x=(1,0,0)$, $\hat{e}_y=(0,1,0)$ and $\hat{e}_z=(0,0,1)$ are unit vectors living in regular, physical space for all $j$, which is necessary in order for $\nabla_{\mathbf{r}_j}\cdot \nabla_{\mathbf{r}_{j'}}\ne 0$ for $j\ne j'$. With this, it is now straightforward to show that $\tilde{\nabla}_S^2$ gives the correct differentiation rule in regular space after projection and, therefore
\begin{equation}
  H_{0,S} = -\frac{\hbar^2}{2m}\tilde{\nabla}_S^2 \equiv \frac{1}{2m}\left(\sum_{j=0}^{n} \mathbf{p}_j\right)^2.\label{superkinetic}
\end{equation}

Regarding the potential $V(\mathbf{r})$, we have essentially absolute freedom of choice when going into superspace. The potential $V_S$ in superspace may be split in any way such that
\begin{equation}
  \lim_{\mathbf{r}_1\to \mathbf{r}}\ldots\lim_{\mathbf{r}_{n}\to \mathbf{r}} V_S(\mathbf{r},\mathbf{r}_1,\ldots,\mathbf{r}_{n}) = V(\mathbf{r}).\label{superpotential}
\end{equation}
It is, however, advisable, to choose a splitting that helps simplify the problem. With all the above considerations, we can write down the full super Hamiltonian $H_S$ as
\begin{equation}
H_S = \frac{1}{2m}\left(\sum_{j=0}^{n}\mathbf{p}_j\right)^2+V_S(\mathbf{r},\mathbf{r}_1,\ldots,\mathbf{r}_{n}).
\end{equation}

A few remarks regarding the general properties of the super Hamiltonian are in order. One should be aware that (i) if the projection of the basis used to solve the problem in superspace onto regular space is overcomplete or linearly dependent, then many of the eigenstates calculated in this way will also be linearly dependent; (ii) the super Hamiltonian admits eigenfunctions whose projections onto regular space vanish identically; (iii) an advantage of working in superspace is that one can choose orthonormal bases whose projections to regular space do not constitute orthogonal -- even if linearly-independent -- basis sets; and (iv) one can construct periodic super Hamiltonians for quasiperiodic Hamiltonians.

\section{Application to a one-dimensional quasiperiodic system}
We devote the rest of the article to demonstrate and showcase the method of section \ref{superspace}, using a realistic one-dimensional model not necessarily in the tight-binding regime consisting of the superposition of two sinusoidal potentials with incommensurate periods.

\subsection{Hamiltonian of the system}
We here study the physics of particles -- atoms or electrons -- in an incommensurate superlattice whose potential $V$ is given by the sum of two periodic potentials $V_1$ and $V_2$ such that
\begin{equation}
  V_i(x+b_i) = V_i(x),\hspace{0.1cm} i=1,2,
\end{equation}
where $b_2/b_1\in \mathbb{R}-\mathbb{Q}$. For concreteness, we choose sinusoidal potentials of the form
\begin{equation}
  V_i(x) = v_i \cos\left(\frac{2\pi x}{b_i}+\phi_i\right),
\end{equation}
where $v_i$ denotes the potential maximum/minimum, and $\phi_i$ are constant phase displacements. The Hamiltonian of the system simply reads
\begin{equation}
  H = -\frac{\hbar^2}{2m}\frac{\partial^2}{\partial x^2} + V_1(x)+V_2(x),\label{Hamiltonian}
\end{equation}
where $m$ is the particle's mass.

\subsection{Solution in superspace}
We consider a two-dimensional superspace, with super Hamiltonian $H_S$ given by (see Eqs.~(\ref{superkinetic}) and (\ref{superpotential}))
\begin{equation}
  H_S = \frac{1}{2m}(p_x+p_y)^2+V_1(x)+V_2(y).
\end{equation}
The original Hamiltonian, Eq.~(\ref{Hamiltonian}), is known to give rise to two types of states at low energies: delocalised and Anderson localised \cite{Frolich}. For simplicity and without loss of generality, we shall consider $v_1=v_2<0$. Then, the ground state will always be delocalised (Anderson localised) for $|v_1|<v_c$ ($|v_1|>v_c$), where $v_c$ is a critical potential strength. The delocalised-localised transition is an example of continuous-to-point spectrum transition \cite{Frolich}. Eigenstates in the continuum are characterised by distributional, discontinuous momentum representations while eigenstates corresponding to the point spectrum have continuous momentum distributions. Since, as we stated earlier, we consider $H_S$ in $L^2(\mathbb{R}^2)$, all its eigenstates satisfy Bloch's theorem with real quasi-momentum $\mathbf{k}$, i.e. if $\psi_{\mathbf{k}}(x,y)$ is an eigenstate of $H_S$ in $L^2(\mathbb{R}^2)$, then
\begin{equation}
  \psi_{\mathbf{k}}(x,y)=e^{i(k_xx+k_yy)}u_{\mathbf{k}}(x,y),\label{psik}
\end{equation}
with real $\mathbf{k}$ and $u_{\mathbf{k}}(x+b_1,y)=u_{\mathbf{k}}(x,y+b_2)=u_{\mathbf{k}}(x,y)$. Eigenstates $\psi(x)$ of $H$ can be generated by projecting the eigenstates of $H_S$, i.e.
\begin{equation}
  \psi(x) = \lim_{y\to x} \psi_{\mathbf{k}}(x,y).
\end{equation}
Since $u_{\mathbf{k}}(x,y)$ is periodic, we can expand it in its Fourier components as
\begin{equation}
  u_{\mathbf{k}}(x,y)=\sum_{n_1,n_2}a_{n_1,n_2}e^{i(\phi_1n_1+\phi_2n_2)}\frac{e^{2\pi i (n_1x/b_1+n_2y/b_2)}}{\sqrt{b_1b_2}}.\label{expansion}
\end{equation}
Introducing the expansion (\ref{expansion}) into Eq.~(\ref{psik}) for $\psi_{\mathbf{k}}$ and into the stationary Schr{\"o}dinger equation $H_S\psi_{\mathbf{k}}=E\psi_{\mathbf{k}}$, we have the following recurrence relation
\begin{align}
  &\frac{\hbar^2}{2m}\left(k_x+k_y+\frac{2\pi n_1}{b_1}+\frac{2\pi n_2}{b_2}\right)^2a_{n_1,n_2}\nonumber\\
  &+\frac{v_1}{2}(a_{n_1+1,n_2}+a_{n_1-1,n_2})+\frac{v_2}{2}(a_{n_1,n_2+1}+a_{n_1,n_2-1})\nonumber\\
  &=Ea_{n_1,n_2}.\label{recurrence}
\end{align}
From the above relation, we immediately observe a degeneracy issue: for any two pairs $(k_x,k_y)$ and $(k_x',k_y')$ such that $k_x+k_y=k_x'+k_y'$, not only is the spectrum identical, but so are their associated eigenfunctions when projected back onto regular space. Therefore, we restrict $(k_x,k_y)$ to $(k_x,0)$. The next observation regarding this issue removes the quasi-momentum from the problem altogether. To see this, define $\ket{k_x}$ as a momentum eigenstate in physical space (one-dimensional). The projected eigenstates have the expansion
\begin{equation}
  \ket{\psi_{(k_x,0)}} \propto \sum_{n_1,n_2}a_{n_1,n_2}\ket{k_x+2\pi (n_1/b_1+n_2/b_2)}.\label{psiexp}
\end{equation}
Since $b_2/b_1$ is an irrational number, the sum above includes (or is dense in) all possible momentum states in the system. In particular, we may choose $k_x' = k_x+2\pi m_1/b_1 + 2\pi m_2/b_2$, with $m_1$ and $m_2$ arbitrary integer numbers. This amounts to a displacement $(n_1,n_2)\to (n_1+m_1,n_2+m_2)$ in Eq.~(\ref{recurrence}), leaving the recurrence relation invariant. It is then easy to see that the projected eigenfunctions and eigenenergies obtained for $k_x'$ and $k_x$ are identical. Therefore, all eigenstates of the original Hamiltonian that can be represented as eigenstates of the super Hamiltonian can be obtained from the latter by choosing $\mathbf{k}=0$, i.e., they are all quasiperiodic functions. 

\begin{figure}[t]
\includegraphics[width=0.5\textwidth]{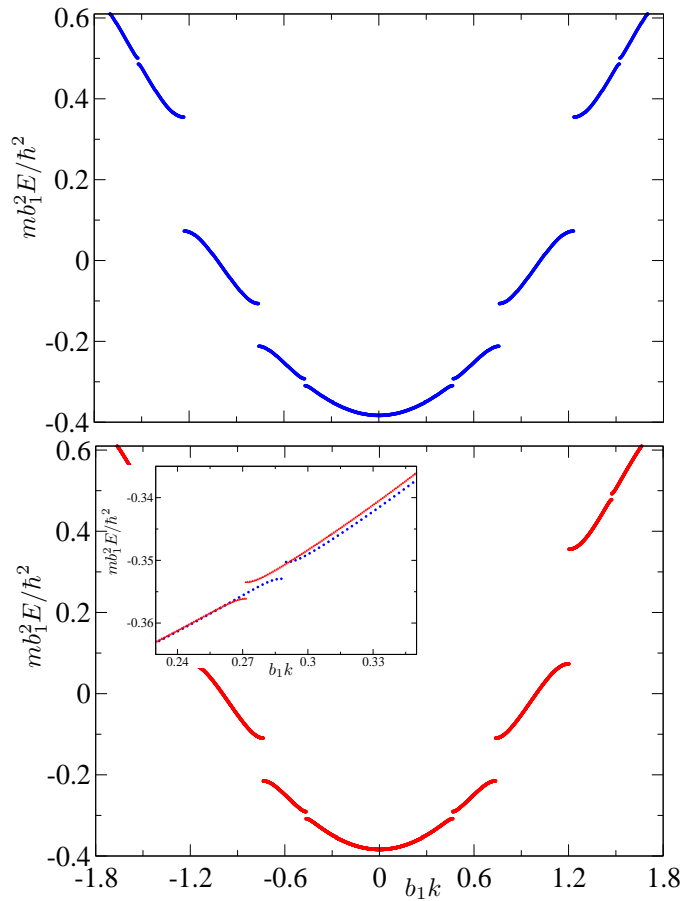}
\caption{Spectrum of the Hamiltonian for $mb_1^2v_1/\hbar^2=mb_1^2v_2/\hbar^2=-2$ in the extended zone scheme (see text), for the quasiperiodic potential (upper row, blue dots) and a periodic approximant with $b_2/b_1=81/50$ (lower row, red dots). The inset is a zoomed in comparison between the spectra.}
\label{fig:bandstructure}
\end{figure}

Numerically, however, we will have to choose a cutoff for $(n_1,n_2)$. With a cutoff, yet well converged eigenfunctions, this actually represents an advantage, since we will be able to easily label states of the system. To see this, choose an integer cutoff $N_c$ such that $|n_1|,|n_2|\le N_c$, and define a momentum scale $k_c>0$ as
\begin{equation}
k_c = 2\pi \times \mathrm{min}\left\{ \left|\frac{n_1}{b_1}+\frac{n_2}{b_2} \right|:\hspace{0.1cm} |n_1|,|n_2|\le N_c,\hspace{0,1cm} (n_1,n_2)\ne \mathbf{0}\right\}\label{kc}
\end{equation}
The scale $k_c$ above represents the state of smallest (in absolute value) non-zero momentum that is included in the expansion of the approximate eigenstate. That is to say that, when a cutoff is used, we do obtain different eigenstates by choosing $-k_c/2 < k_x \le k_c/2$ in the low-energy manifold \footnote{Since $k_c$ is the smallest momentum, but not the smallest momentum difference that can be achieved, at higher energies $k$ may be restricted to a shorter interval.}. These states, although quasi-momentum has no physical meaning, constitute a continuous set whose associated spectrum can be easily visualised either in this artificially constructed ``Brillouin zone'', or in the extended zone scheme, which is independent of the choice of cutoff (provided convergence has been achieved). The algorithm we use to extract the spectrum is as follows: (i) we choose $N_c$ and identify $k_c$, Eq.~(\ref{kc}); (ii) we solve the recurrence relation (\ref{recurrence}) and obtain all eigenvalues for $k\in (-k_c/2,k_c/2]$, with a small step $\Delta k$ between different $k$'s; (iii) when a gap opens at $|k|<k_c/2$ there are more than two degenerate states (not allowed by symmetry), and we discard those states with the worst local energy $E(x)=H\psi(x)/\psi(x)$; (iv) we increase $N_c$ and go back to (i) until convergence is achieved; and (v) we order the states in increasing (decreasing) energy for $k\ge 0$ ($k<0$) in steps of $\Delta k$ (extended zone scheme). In all calculations we consider $b_2/b_1 = (1+\sqrt{5})/2$, i.e. the golden ratio, and set $\phi_1=\phi_2=0$ for concreteness, and without loss of generality. In Fig.~\ref{fig:bandstructure} we plot the spectrum obtained in this way for $mb_1^2v_1/\hbar^2=-2$ in the extended zone scheme, together with the band structure for a periodic approximant with $b_2/b_1 = 81/50$ in the extended zone scheme. Overall, they are in good agreement. However, with a periodic approximant band gaps can only be located at the band centre ($k=0$) and edges ($k=\pm \pi/81$) in the Brillouin zone. In the inset of Fig.~\ref{fig:bandstructure}, we present a zoomed-in portion of the spectrum, where it is clearly observed that both the gap locations and sizes are different. For Anderson localised eigenstates, which by definition correspond to point spectrum, convergence using this method is never achieved due to the fact that the momentum distribution of a bound state is continuous (compare with Eq.~(\ref{psiexp})). As a result, we shall follow a different route.

\subsection{Defects and Density of States.} One major consequence of the continuous spectrum in the delocalised regime, is the possibility of writing down scattering states off impurities. These are important for the characterisation of quantum transport in incommensurate lattices. We now calculate the single-particle Green's function exactly and then use it to obtain the scattering properties in the system, and the local density of states (LDOS) and the density of states (DOS).

The Green's function $G_0(E;x,x')$ satisfies the following equation
\begin{equation}
  (E-H_0(x))G_0(E;x,x')=\delta(x-x'),\label{GreenFunctionEquation}
\end{equation}
with $H_0$ given by $H$ in Eq.~(\ref{Hamiltonian}). We set physical boundary conditions for the Green's function so that $G_0(E;x,x')\sim \psi_k(x)$, if the group velocity ($(1/\hbar)\mathrm{d}E(k)/\mathrm{d}k$) associated with $\psi_k$ is positive, and $G_0(E;x,x')\sim \psi_{-k}(x)$ otherwise -- which is relevant only if we work in the reduced zone scheme. We shall work here in the extended zone scheme, where $k$ labels monotonically increasing energy states. It is easy to see that the Green's function must take the form
\begin{align}
  G_0(E;x,x') &= \theta(x-x')A_k(x')\psi_k(x)\nonumber \\
  &+\bar{\theta}(x'-x)A_{-k}(x')\psi_{-k}(x),\label{GreenFunctionForm}
\end{align}
where $\theta(x)$ ($\bar{\theta}(x)$) is the Heaviside step function being zero (unity) at $x=0$, and the functions $A_k(x)$ and $A_{-k}(x)$ remain to be determined. The diagonal of the Green's function must exist, from what we find $A_k(x)\psi_k(x)=A_{-k}(x)\psi_{-k}(x)$. We introduce this relation, together with Eq.~(\ref{GreenFunctionForm}) into Eq.~(\ref{GreenFunctionEquation}) and obtain the following expression for $A_k$
\begin{equation}
A_k(x)=\frac{(2m/\hbar^2)\psi_{-k}(x)}{\psi_{-k}(x)\partial_x\psi_k(x)-\psi_{k}(x)\partial_x\psi_{-k}(x)}.\label{Ak}
\end{equation}
Calculating scattering states $\Psi_k$ off static impurities with potential $W(x)$ now reduces to solving the Lippmann-Schwinger equation 
\begin{equation}
  \Psi_k(x) = \psi_k(x) + \int_{-\infty}^{\infty}\mathrm{d}x'G_0(E;x,x')W(x')\Psi_k(x').
\end{equation}
The simplest type of impurity that can be studied this way is a zero-range impurity of strength $g$ located at $x=x_0$, for which $W(x)=g\delta(x-x_0)$. This problem is exactly solvable and gives for the transmission ($t$) and reflection ($r$) coefficients
\begin{align}
r&=\frac{g \psi_k(x_0)}{1-gA_k(x_0)\psi_k(x_0)}A_{-k}(x_0),\label{reflection}\\
t&=\frac{g \psi_k(x_0)}{1-gA_k(x_0)\psi_k(x_0)}A_{k}(x_0)+1\label{transmission}.
\end{align}
Clearly, since all quantities involved in Eqs.~(\ref{reflection}) and (\ref{transmission}) are quasiperiodic, both $r$ and $t$ are quasiperiodic functions of the impurity's position $x_0$ for fixed $g$.

\begin{figure}[t]
\includegraphics[width=0.5\textwidth]{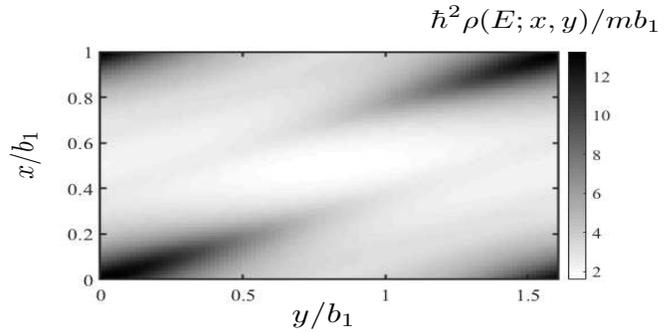}
\caption{Local super density of states for $mb_1^2v_1/\hbar^2=mb_1^2v_2/\hbar^2=-2$ and energy $mb_1^2E/\hbar^2=-0.38227376$.}
\label{fig:LSDOS}
\end{figure}

The local DOS $\rho(E,x)$ is obtained from the diagonal part of the Green's function as $\rho(E,x) = -(1/\pi)\mathrm{Im}G_0(E;x,x)$. Using Eqs.~(\ref{GreenFunctionForm}) and (\ref{Ak}), and choosing the gauge where $\psi_k^*=\psi_{-k}$, it simplifies to
\begin{equation}
  \rho(E;x) = \frac{m}{\pi\hbar^2}\frac{|\psi_k(x)|^2}{\mathrm{Im}\left[\psi_k^*(x)\partial_x\psi_k(x)\right]}\label{LDOS}
\end{equation}
Notice that, in the free particle case, Eq.~(\ref{LDOS}) reduces to $\rho(E,x) = m/\pi\hbar^2k$, as is well known, and that all the relations above apply equally well to a particle in an arbitrary periodic potential, or any other system that supports continuous spectrum (including scattering states off impurities). Moreover, identifying the particle density $\mathcal{R}_E(x)=|\psi_k(x)|^2$ and current $j_E(x)=\hbar^2\mathrm{Im}(\psi^*\partial_x\psi)$ in Eq.~(\ref{LDOS}), we have an expression for the LDOS in terms of gauge-invariant quantities
\begin{equation}
  \rho(E;x)=\frac{1}{\pi}\frac{\mathcal{R}_E(x)}{\left|j_E(x)\right|}=\frac{1}{\pi |v_E(x)|},
\end{equation}
where we have further identified $j_E = \mathcal{R}_Ev_E$, with $v_E(x)$ the local velocity.

\begin{figure}[t]
\includegraphics[width=0.5\textwidth]{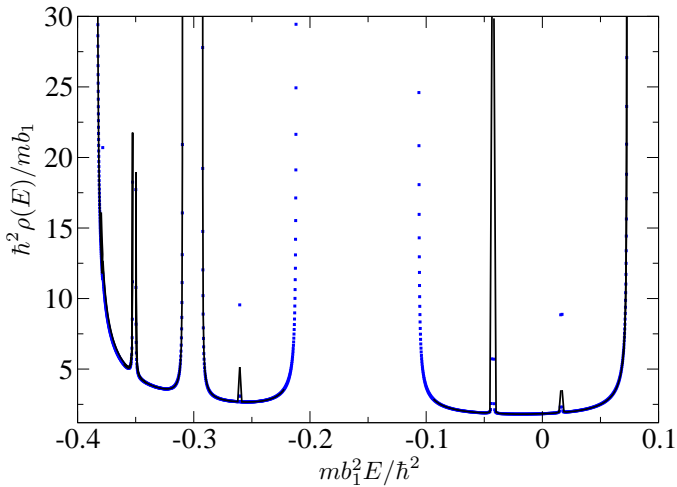}
\caption{Density of states for $mb_1^2v_1/\hbar^2=mb_1^2v_2/\hbar^2=-2$ as a function of the energy, calculated using Eq.~(\ref{calculaterho}) (solid black lines), and the energy spectrum (blue dots).}
\label{fig:DOS}
\end{figure}

For the DOS $\rho(E)$, due to the incommensurability of the problem, we must calculate
\begin{equation}
  \rho(E) = \lim_{L\to \infty}\int_{-L/2}^{L/2}\frac{\mathrm{d}x}{L}\rho(E,x).\label{rhoE1}
\end{equation}
The calculation of $\rho(E)$ above can be simplified by noticing that the local density of states, Eq.~(\ref{LDOS}), is a quasiperiodic function. To see this, assume that $\psi_k$ has been calculated with the help of a momentum scale $q\in(-k_c/2,k_c/2]$ as explained above ($q\ne k$, since $k$ is simply a label), so that $\psi_k(x)=\exp(iqx)\phi(x)$, with $\phi(x)$ quasiperiodic. We define the local super density of states LSDOS $\rho^{\mathrm{S}}(E;x,y)$ as the extension of Eq.~(\ref{LDOS}) to superspace, i.e. so that
\begin{equation}
\rho(E;x) = \lim_{y\to x} \rho^{\mathrm{S}}(E;x,y).\label{correspondence}
\end{equation}
This is achieved by using the wave functions $\psi_k(x,y)$ calculated in superspace and recalling the rule $\partial_x\to \partial_x+\partial_y$. After trivial algebraic manipulations, we obtain
\begin{equation}
\rho^{\mathrm{S}}(E;x,y) = \frac{m}{\pi \hbar^2q} \left[1+\frac{\mathrm{Im}\left[\phi^*(x,y)(\partial_x+\partial_y)\phi(x,y)\right]}{q\left|\phi(x,y)\right|^2}\right]^{-1},
\end{equation}
which is clearly a periodic function, a particular case of which we plot in Fig.~\ref{fig:LSDOS}. The LSDOS $\rho^{\mathrm{S}}(E;x,y)$ therefore admits a Fourier series as
\begin{equation}
\rho^{\mathrm{S}}(E;x,y)=\sum_{n_1,n_2}\rho_{n_1,n_2}\frac{e^{i2\pi \left(\frac{n_1}{b_1}x+\frac{n_2}{b_2}y\right)}}{\sqrt{b_1b_2}}.
\end{equation}
Using the above equation, together with Eqs.~(\ref{correspondence}) and (\ref{rhoE1}), we obtain
\begin{equation}
\rho(E) = \frac{\rho_{0,0}}{\sqrt{b_1b_2}} \equiv \frac{1}{b_1b_2}\int_{0}^{b_1}\mathrm{d}x\int_{0}^{b_2}\mathrm{d}y\rho^{\mathrm{S}}(E;x,y).\label{calculaterho}
\end{equation}
In Fig.~\ref{fig:DOS} we plot the DOS for $mb_1^2v_1/\hbar^2=mb_1^2v_2/\hbar^2=-2$ using both the energy spectrum and Eq.~(\ref{calculaterho}), finding excellent agreement between the two. This shows that the label $k$, with dimensions of momentum, is a well-defined quantity exactly corresponding to the so-called rotation number used in Mathematical Physics \cite{MoserRotationNumber,AvronSimon}.

\subsection{Localised states.} As the strength of the quasiperiodic potential is increased, it is known that states in the lowest-energy portion of the spectrum become Anderson localised \cite{Frolich}. By definition, these states correspond to point spectrum, for which the momentum distribution is continuous. Therefore, we cannot expect the discrete expansion (\ref{expansion}) to converge and we have verified it is impossible to numerically converge with a finite basis in the localised regime. These statements are not in contradiction with the quasiperiodicity of the solutions stemming from Bloch's theorem in superspace, provided we allow distributional solutions (e.g. Dirac delta functions) to the problem. We build upon this in Appendix \ref{apendicitis}.

For the numerical calculations, we must give up the infinite-size limit in favour of large, yet finite boxes in position space. We begin with the fundamental recurrence relation, Eq.~(\ref{recurrence}). We define new ``coordinates'' as $R=n_1/b_1+n_2/b_2$ and $n=n_1-n_2$. After setting $\mathbf{k}=0$, the recurrence relation (\ref{recurrence}) becomes
\begin{align}
  &\left[\frac{\hbar^2}{2m}(2\pi R)^2-E\right]a_{R,n}\nonumber\\
  &+\frac{v_1}{2}(a_{R+1/b_1,n+1}+a_{R-1/b_1,n-1})\nonumber\\
  &+\frac{v_2}{2}(a_{R+1/b_2,n-1}+a_{R-1/b_2,n+1})=0.
\end{align}
Since the diagonal part of the above recurrence relation does not depend on $n$, the problem is separable as $a_{R,n}=\Phi(R)f_n$, and $f_n=\exp(i\lambda n)$ are solutions. Noting that we are working in superspace, it is straightforward to see that only the solution with $\lambda=0$ is non-vanishing for $y\to x$ and we therefore choose this solution. Note that $R$ can take on any real value. This can be used to our advantage since, for continuous $\Phi(R)$, we can use the scalar product on $L^2(\mathbb{R})$. In order to solve the problem numerically, we choose a plane wave basis, and a momentum scale $\kappa$, and impose periodic boundary conditions on $\Phi(R)$ in $(-\kappa/2,\kappa/2]$. The members $\phi_{\ell}$ of the plane-wave basis are given by $\phi_{\ell}(R) = \exp(2\pi i \ell R/\kappa)/\sqrt{\kappa}$. Clearly, for $\kappa\to \infty$ the problem is simply the original problem in the momentum representation. It is however very convenient to consider $\kappa$ as small as possible, as long as convergence is achieved. This can be seen by writing $\Phi(R)\approx\sum_{\ell=-N_c}^{N_c}\alpha_{\ell} \phi_{\ell}(R)$, with $N_c$ an integer cutoff required for the calculation. Then the real-space wave function reads
\begin{equation}
  \psi(x)\propto \sum_{\ell=-N_c}^{N_c}\alpha_{\ell} \frac{\sin\left[2\pi\left(\frac{\ell}{\kappa}+x\right)\frac{\kappa}{2}\right]}{\pi\left(\frac{\ell}{\kappa}+x\right)},
\end{equation}
which vanishes for $|x|=(N_c+1)/\kappa$, effectively setting open boundary conditions at these points. We remark that this method also works to implement open boundary conditions in the delocalised regime. This method has the great appeal that, once convergence has been achieved for fixed values of $\kappa$ and $N_c$, increasing $N_c$ increases the size of the box ($L=2(N_c+1)/\kappa$) while retaining convergence.

\begin{figure}[t]
\includegraphics[width=0.5\textwidth]{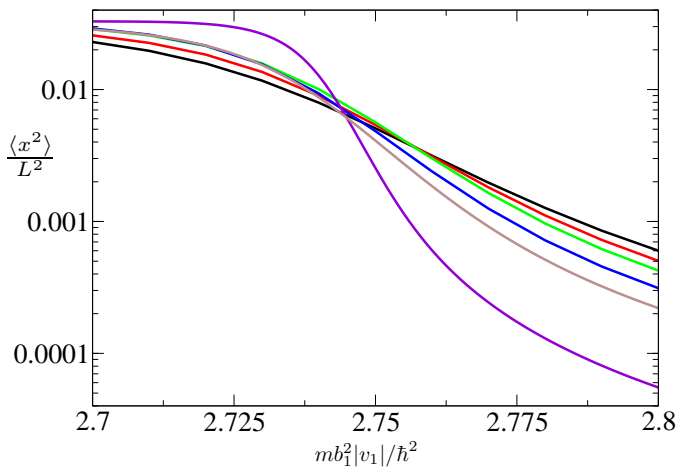}
\caption{Extent of the ground state wave function in a finite box of size $L$ as a function of $v_1=v_2<0$. On the right end, bottom to top curves correspond to $L/b_1 =$ $666$, $333$, $280$, $240$, $220$ and $200$.}
\label{fig:x2}
\end{figure}

A marker of localisation is given by the variance of the position \cite{Nathan} which, for the ground state, which is spatially symmetric, coincides with $\langle x^2 \rangle$. Since it is known that, for the model we are using here as an example, localisation occurs first in the ground state after a critical value for $|v_1|$ is reached \cite{Frolich}, also recently confirmed numerically by Sanchez-Palencia {\it et al.} \cite{SanchezPalencia}, we focus solely on the ground state. There are three different possibilities, namely that this is delocalised, localised or critical. If it is delocalised, then $\langle x^2 \rangle / L^2 \to C$, with $C>0$ a constant that coincides with the value obtained for a free particle, as $L\to \infty$. If it is localised, then obviously $\langle x^2 \rangle / L^2 \to 0$ for large $L$. At the critical point, we expect scale invariance, i.e. the ground state wave function behaving as a power law at long distances and, therefore, $\langle x^2 \rangle/L^2 \to C'$, with $C'\ne C$ a constant and, moreover, its extent should exhibit, for large enough system sizes, negligible finite-size effects. That is to say, that for large $L$, all curves $\langle x^2 \rangle/L^2$ should cross at the critical point. In Fig.~\ref{fig:x2} we plot the extent of the ground state wave function for increasingly large $L$ ($L/b_1=200$ to $L/b_1=660$), and observe that indeed the curves cross at approximately the same point, which we can estimate as $mb_1^2|v_1|/\hbar^2\approx 2.74$. An analysis of the effective mass in the delocalised regime will show that this is indeed the case.  

\subsection{Delocalised-localised transition.} 
\begin{figure}[t]
\includegraphics[width=0.5\textwidth]{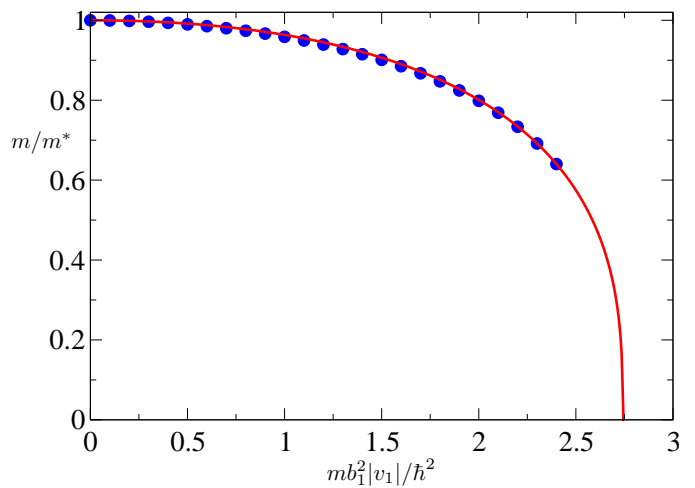}
\caption{Inverse effective mass near the ground state as a function of the potential strength for $v_1=v_2<0$ (blue dots). The red solid line is the fit of Eq.~(\ref{curve}).}
\label{fig:effectivemass}
\end{figure}

The transition from delocalised quasiperiodic Bloch states to Anderson localised states, near the ground state, can be easily studied from the delocalised side of the problem where we have very detailed information about the infinite-size spectrum. A quantity that reflects the transition in the ground state is its effective mass. The effective mass for a discrete state is infinite -- it is localised -- while it is finite for continuum states. From the energy dispersion $E=E(k)$ we have for the effective mass $m^*$
\begin{equation}
 \lambda \equiv \frac{m}{m^*} = \frac{m}{\hbar^2}\frac{\mathrm{d}^2E(k)}{\mathrm{d}k^2}|_{k=0}.
\end{equation}
Near the critical point $v_c$, we expect the effective mass to be scale invariant, i.e. $\lambda = \lambda(v_1)$ is approximately
\begin{equation}
  \lambda(v_1)\approx \alpha \left(1-\frac{|v_1|}{v_c}\right)^{\gamma},\label{ansatz1}
\end{equation}
for $|v_1|\to v_c^-$, where $\alpha$ is a numerical constant. As we approach the critical point, however, convergence becomes more difficult to achieve. We have verified that, for a cutoff $N_c=20$, our numerics converge well up until $|v_1|=2.45$. From the analysis of the localised side of the problem, we have estimated that $v_c\approx 2.74$. In order to obtain a better fit to the critical behaviour of $\lambda$ further away from the critical point, we modify the function in Eq.~(\ref{ansatz1}) such that (i) it is given by Eq.~(\ref{ansatz1}) very close to the critical point; (ii) $\lambda(0)=1$; and (iii) it is quadratic in $|v_1|$ at low $|v_1|$, which is verified in our problem. The simplest function satisfying conditions (i -- iii) has the form
\begin{equation}
  \lambda(v_1) \approx \left(1-\frac{|v_1|}{v_c}\right)^{\gamma}\left(1+\gamma\frac{|v_1|}{v_c}\right).\label{curve}
\end{equation}
We choose a small interval $|v_1|\in [2.3,2.45]$ to perform a fit to $m/m^*$, using a fine spacing between different values of $|v_1|$ obtaining $mb_1^2v_c/\hbar^2=2.7411\pm 10^{-4}$ and $\gamma = 0.33861\pm 5\cdot 10^{-5}$. The critical point is essentially exact, as this was accurately computed in Ref.~\cite{SanchezPalencia} as $mb_1^2v_c/\hbar^2=2.7410$. The value of the critical exponent is consistent with $\gamma=1/3$. Remarkably, the scaling of the finite-size effects in the inverse participation ratio, from the localised regime, exhibits also a critical exponent that is consistent with the value $1/3$ \cite{SanchezPalencia}. The curve (\ref{curve}) is plotted in Fig.~(\ref{fig:effectivemass}), alongside the calculated values of $m/m^*$. Note how the ansatz (\ref{curve}) nicely interpolates between low $|v_1|$ and the critical point. The critical behaviour in Eq.~(\ref{ansatz1}) is therefore realised for $\alpha = 1+\gamma$.  

\subsection{Edge states}
As is well-known, Bloch waves of periodic (i.e. commensurate) superlattices can have non-trivial topological properties \cite{Cooper}. It is reasonable, from continuity arguments, to assume that quasiperiodic superlattices may hold the same non-trivial topological properties as their periodic counterparts. This reasoning was used in Ref.~\cite{Zilberberg} to argue that this may be the case in a tight-binding model using periodic approximants. This also carries the problem of having to deal with ever smaller Brillouin zones as better periodic approximants are used. However, a definitive proof requires to avoid the departure from full quasiperiodicity. There are a number of ways in which this could be done: (i) by means of the Bott index \cite{HuangLiu}, which does not require periodic Brillouin zones; (ii) by using the extended zone scheme to compute the Zak phase \cite{Zak}(or change thereof from one ``band'' to the next), integrating over $(\alpha,\beta]\cup(-\beta,-\alpha]$, where $\alpha$, $\beta$ $>0$ are two consecutive band gap locations; or (iii) by showing there exist in-gap, non-normalisable states with an infinite number of nodes which are either exponentially growing or decaying as $x\to \infty$ \cite{DuncanOhbergValiente}.      
\begin{figure}[t]
\includegraphics[width=0.5\textwidth]{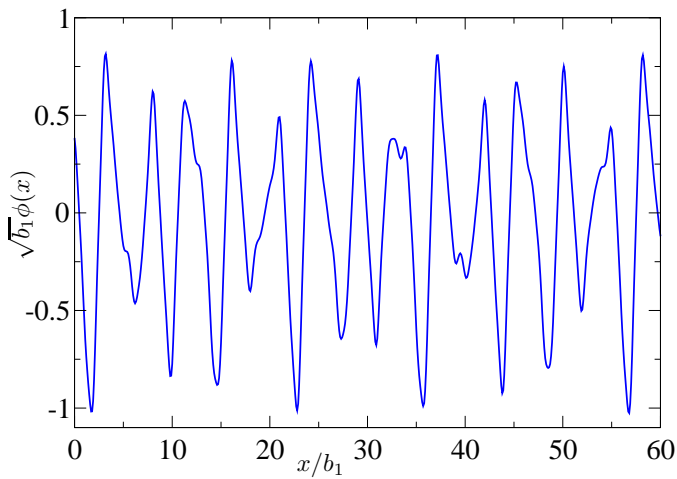}
\caption{Portion of the quasiperiodic part of an edge state for $mb_1^2v_1/\hbar^2=mb_1^2v_2/\hbar^2=-2$ (see text), with energy $mb_1^2E/\hbar^2\approx 0.4909905$, which is an eigenstate on every semi-infinite line for $x>x_0$, with $x_0$ any of its nodes.}
\label{fig:edgestate}
\end{figure}
We shall not deal with option (i) above, which is more complicated and is not required here. We note that the Bott index is a very powerful technique to deal with true random disorder (see Ref. \cite{HuangLiu} for details). The Zak phase $\mathcal{Z}$ in the quasiperiodic case is calculated as     
\begin{equation}
  \frac{\mathcal{Z}}{\beta-\alpha}=i\int_{-\beta}^{-\alpha}\mathrm{d}k\langle \psi_k \left|\right.\partial_k\psi_k\rangle +i\int_{\alpha}^{\beta}\mathrm{d}k\langle \psi_k \left|\right.\partial_k\psi_k\rangle ,
\end{equation}
where $\psi_k$ is a solution corresponding to the momentum label $k$. The calculation of the Zak phase, even though {\it a priori} possible, carries the difficulty of identifying some of the band gaps, which can be extremely small in magnitude, with a finite numerical resolution. We therefore take the third option, which is by far the simplest. To calculate the edge modes, we allow the momentum $k$ in Eq.~(\ref{recurrence}) to take on complex values, and look for real eigenvalues. For $\mathrm{Im}(k) > 0$ ($<0$), the solutions are quasiperiodic functions times a decaying (increasing) exponential, $\phi(x)\exp(ikx)$, as $x\to \infty$, corresponding to edge modes in semi-infinite space bound to the right (left) of an infinite wall placed at an arbitrary position $x_0$ that coincides with a node of the quasiperiodic function. We have found edge modes of this type (not shown) in every band gap we have identified for $mb_1^2v_1/\hbar^2 = -2$ within the energy ranges shown in Fig.~\ref{fig:bandstructure}, spanning all energies within the gaps by changing the values of $k$, which effectively changes the position of, say, the first positive node of the edge mode wave function, or equivalently the relative phase between the two periodic potentials in Hamiltonian~(\ref{Hamiltonian}) while keeping the position of the wall fixed. In Fig.~\ref{fig:edgestate}, we plot a portion of the quasiperiodic function $\phi(x)$ corresponding to an edge mode with energy $mb_1^2E/\hbar^2\approx 0.4909905$, for which $b_1k=10^{-2}i$, and extends to infinity.  

\section{Conclusions and outlook}
We have introduced a Hamiltonian formalism that extends the dimension of space to what is known as superspace, with quasiperiodic potentials and quasicrystals in mind. We have shown that it is necessary to proceed with care since the extended ``super'' Hamiltonian is quite degenerate, and have provided ways to overcome this difficulty successfully. We have applied our theory to a one-dimensional particle in a quasiperiodic potential and obtained a number of results and found that the superspace approach is most useful for continuous spectra. We have also obtained the most general single-particle Green's function in one dimension which is valid for any one-dimensional single-particle system in its continuum, and obtained the density of states and related quantities from it. We have shown that scattering states off impurities or defects can be easily calculated using this Green's function, which is of relevance for quantum transport in non-trivial media, and showed how to extract semi-infinite topological edge states directly from the superspace formalism.

The work here presented constitutes a proof of principle regarding dimensional extensions to solve quantum mechanical problems. As such, there are many open problems and a large number of possible improvements. It is worth mentioning that the superspace method is not restricted to quasiperiodic systems. It may also be used to bypass non-orthogonality, and even overcompleteness, within a natural basis for a quantum problem. A simple example we can think of is that of zero-range-interacting one-dimensional bosons in a harmonic trap. Indeed, a natural basis would comprise the eigenstates of the harmonic oscillator multiplied by homogeneous few-body scattering states, which are exactly solvable via the Bethe ansatz \cite{LiebLiniger}. A basis of this type is obviously not orthogonal, and extension to superspace may prove useful. Two other important, interconnected ideas that are yet to be explored concern quasiperiodic and quasicrystalline systems. Firstly, for states in the continuum convergence and quality of results may be vastly improved upon by ``re-diagonalising'' the system. That is, by taking a number of unphysically-degenerate (or near-degenerate), yet non-identical states calculated in a finite basis in superspace, where these are orthogonal, and using them as a small non-orthogonal basis in physical space to obtain better results. Secondly, one may take far-from-converged states obtained via superspace after the localisation transition and construct localised Wannier-type orbitals to, again, re-diagonalise the system in physical space (see appendix \ref{apendicitis}). Our studies can be used in higher dimensions, just at a higher computational cost, and should be generalisable to tight-binding quasicrystals, which may be simpler from a computational point of view. Research is on-going in this direction.

\begin{acknowledgments}
MV is indebted to Chris J. Pickard and Bartomeu Monserrat for introducing the concept of superspace to him and for illuminating discussions about incommensurate systems. We thank Xiao-Tian Nie for invaluable input regarding notation in an earlier draft of this manuscript. CWD acknowledges support from EPSRC CM-CDT Grant No. EP/L015110/1, and NTZ by a Jens Chr. Skou fellowship from the Aarhus University Research Foundation.
\end{acknowledgments}

\appendix

\section{Quasiperiodicity of localised eigenstates}\label{apendicitis}
Here we show that, if distributional solutions to the eigenvalue problem in superspace are allowed, then localised eigenstates can be defined as singular quasiperiodic solutions. Firstly, we can show that for strong quasiperiodic potentials, the eigenstates must be localised. To see this, set $v_1<0$ and $v_2<0$ in Eq.~(\ref{recurrence}), with $\mathbf{k}=0$. This is not a restriction as we are free to choose the phase of the potentials, but only a way to access the ground state easily. For $|v_1|$ and $|v_2|\to \infty$, since $(n_1/b_1+n_2/b_2)^2$ is bounded from below, we have the strong-coupling limit of the (unnormalised) ground state
\begin{equation}
a_{n_1,n_2}=1, \hspace{0.1cm} \forall n_1,\hspace{0.01cm} n_2\label{a1}
\end{equation}
with energy $E=v_1+v_2$. The eigenfunction in superspace (assuming $\phi_1=\phi_2=0$ for simplicity) reads
\begin{align}
  \psi(x,y)&\propto \sum_{n_1,n_2}e^{2\pi i (n_1x/b_1+n_2y/b_2)} \nonumber \\
  &\propto \sum_{m_1,m_2}\delta(x-m_1b_1)\delta(y-m_2b_2).\label{eigen1}
\end{align}
Clearly, the only non-zero contribution in the second line of Eq.~(\ref{eigen1}) to the projection onto physical space corresponds to $m_1=m_2=0$, which gives $\psi(x) \propto \delta(x)$, because $b_2/b_1$ is irrational. This is to be expected in the strong-coupling limit, since the Dirac delta function (properly regularised) is the only possible positive distribution that is infinitely narrow. Since $\psi$ is localised, it corresponds to point spectrum. This shows that there exists a critical point in the space of $(v_1,v_2)$ for irrational $b_2/b_1$ that takes the ground state associated with continuous (delocalised) spectrum to point (localised) spectrum. Note that, were the original Hamiltonian periodic, i.e. $b_2/b_1$ rational, Eq.~(\ref{a1}) would still be valid but $\psi(x)$ would be periodic (extended) and, therefore, corresponding to continuous spectrum, as is well known from Bloch's theorem.

We now address the problem for localised eigenfunctions at finite values of the potential strengths. In this case, $a_{n_1,n_2}$ is not a constant, but instead we have
\begin{equation}
  a_{n_1,n_2}=\int_{-\infty}^{\infty}\mathrm{d}\lambda f(\lambda) e^{2\pi i(n_1/b_1+n_2/b_2)\lambda}.
\end{equation}
The eigenfunction in superspace $\psi(x,y)$ takes the form
\begin{equation}
  \psi(x,y) \propto \int \mathrm{d}\lambda f(\lambda) \sum_{m_1,m_2}\delta(x-\lambda-m_1b_1)\delta(y-\lambda-m_2b_2),\label{Wannier}
\end{equation}
and, therefore $\psi(x,y)\propto \delta(x-y)f(x)$, from which we see that $f(x)$ is nothing but the eigenfunction in physical space. Note how, for commensurate superlattices, $f(x)$ is actually a Wannier function. This suggests the possibility of constructing new basis sets for incommensurate superlattices using the Wannier functions associated with non-converged quasiperiodic Bloch waves in superspace in the localised regime, which could be very helpful in reducing finite-size effects and in higher dimensions.

\section{Examples with functions not of the product form}\label{apendicitis2}
Here, we illustrate, by using two different superspace representations of a wave function in $L^2(\mathbb{R})$, that the definition of the gradient in superspace, Eq.~(\ref{superspacegradient}), is the correct one. Consider the following wave function $\psi(x)$
\begin{equation}
  \psi(x) = e^{-\sqrt{a^2+x^2}}.\label{psixapp2}
\end{equation}
Its partial derivative with respect to $x$ is just
\begin{equation}
  \partial_x\psi(x) = -\frac{x}{\sqrt{a^2+x^2}}e^{-\sqrt{a^2+x^2}}.
\end{equation}
We choose a three-dimensional superspace, with wave functions in $L^2(\mathbb{R}^3)$, and two particular examples, $\Psi_1(x,y,z)$ and $\Psi_2(x,y,z)$, for which it holds that
\begin{equation}
  \lim_{z\to x}\lim_{y\to x}\Psi_1(x,y,z)= \lim_{z\to x}\lim_{y\to x}\Psi_2(x,y,z)=
  \psi(x),\label{limitapp2}
\end{equation}
with $\psi(x)$ given by Eq.~(\ref{psixapp2}). These two functions are chosen as
\begin{align}
  \Psi_1(x,y,z)&=e^{-\sqrt{a^2+(x^2+y^2+z^2)/3}},\\
  \Psi_2(x,y,z)&=e^{-\sqrt{a^2+(x+y-z)^2}},
\end{align}
so that Eq.~(\ref{limitapp2}) clearly holds. We apply the gradient in superspace, Eq.~(\ref{superspacegradient}), on $\Psi_1$ and $\Psi_2$, obtaining
\begin{align}
  (\partial_x+\partial_y+\partial_z)\Psi_1(x,y,z)&=-\frac{x+y+z}{3\sqrt{a^2+(x^2+y^2+z^2)/3}}\nonumber \\
  &\times e^{-\sqrt{a^2+(x^2+y^2+z^2)/3}}.\\
  (\partial_x+\partial_y+\partial_z)\Psi_2(x,y,z)&=-\frac{x+y-z}{\sqrt{a^2+(x+y-z)^2}}\nonumber\\
  &\times e^{-\sqrt{a^2+(x+y-z)^2}}.
 \end{align}
From the above equations, it is clear that
\begin{align}
  \lim_{z\to x}\lim_{y\to x}\left[(\partial_x+\partial_y+\partial_z)\Psi_1(x,y,z)\right]&=\nonumber\\
  \lim_{z\to x}\lim_{y\to x}\left[(\partial_x+\partial_y+\partial_z)\Psi_2(x,y,z)\right]&=\partial_x\psi(x),
\end{align}
as we wanted to show.

\bibliographystyle{unsrt}

\end{document}